\documentclass[preprint,prl,amsmath,amssymb]{revtex4}

\usepackage{graphicx}
\usepackage{dcolumn}
\usepackage{bm}

\begin{document}

\title{Quantum many particle systems in ring-shaped optical lattices}

\author{Luigi Amico$^{(a)}$, Andreas Osterloh$^{(b)}$,
and Francesco Cataliotti$^{(c)}$}

\affiliation{CRS MATIS-INFM, via S. Sofia 64, 95125 Catania, ITALY}

\affiliation{(a)\, Dipartimento di Metodologie Fisiche e
    Chimiche (DMFCI), Universit\'a di Catania, viale A. Doria 6,
95125 Catania, ITALY}

\affiliation{(b)\, Institut f\"ur Theoretische Physik, Universit\"at Hannover,  30167 Hannover, GERMANY}

\affiliation{(c)\, LNS-INFN $\&$ Dipartimento di  Fisica $\&$
Scuola Superiore di Catania, Universit\'a di Catania, via S. Sofia
64, 95125 Catania, ITALY}

\begin{abstract}
In the
present work we demonstrate how to realize 1d-optical closed
lattice experimentally, including a {\it tunable} boundary
phase-twist.
The latter may induce  ``persistent currents'',
visible by studing the atoms' momentum distribution.
We show how important phenomena in 1d-physics can be studied by physical
realization of systems of trapped atoms in ring-shaped optical lattices.
A mixture of bosonic and/or fermionic atoms can be loaded into the lattice,
realizing a generic quantum system of many interacting particles.
\end{abstract}

\maketitle

Studies of one dimensional systems constitute an intense research
activity both in experimental and theoretical physics. They are
particularly interesting mainly because quantum effects are
strongest at low dimensionality and peculiar phenomena emerge.
Prominent examples are the spin-charge separation in Luttinger
liquids\cite{LL}, one dimensional persistent currents in
mesoscopic rings\cite{PERSISTENT}, and transmutation of quantum
statistics\cite{FRACTIONAL}. Most of the approximate schemes
working in higher dimensions break down in 1-d. Only for a
restricted class of model Hamiltonians, physical properties can be
obtained analytically resorting to powerful techniques as Bethe
ansatz\cite{TAKAHASHI} or conformal field theory\cite{CONFORMAL}.
For more generic 1d systems, numerical analysis is the standard
route to extract physical information. Degenerate atoms in optical
lattices could constitute a further tool for the
investigations\cite{CIRAC}, thus rediscovering  Feynman's
ideas\cite{FEYNMAN}  suggesting that an ideal system with a
``quantum logic'' can be used to study open problems in quantum
physics. Precise knowledge of the model Hamiltonian, manipulation
of its coupling constants, possibility of working with
controllable disorder are some of the great advantages of atomic
systems in optical lattices compared with solid state devices
 to experimentally realize Feynman's ideas.
The upsurge of interest of
the scientific community has been remarkable,  and some
perspectives disclosed by trapped--atom ``labs'' have been
already explored: the observation of the superfluid--Mott
insulator quantum phase transition\cite{GREINER}, the analysis of
the Tonks-Girardeau regime in strongly interacting
bosons\cite{PAREDES}, and the physical realization of a $1d$-chain
of Josephson junctions\cite{CATALIOTTI} were  relevant
achievements for condensed  matter physics.
The two most widely used methods to trap and manipulate atoms are
based on the conservative interaction of atoms with either
magnetic fields or with far off--resonant laser beams. For our
purposes the magnetic trapping potential has a parabolic symmetry.
Laser light interacts with the atomic induced dipoles creating
attractive or repulsive potentials depending on the sign of the
detuning $\Delta$ from resonance~\cite{Grimm}.
This can be used to create different  potentials for different
atoms, but with a single tunable laser beam. Notice  that no light
absorption occurs in  creating  the potential; therefore the medium can be
considered  transparent to the laser.

\begin{figure}
\includegraphics[width=6cm]{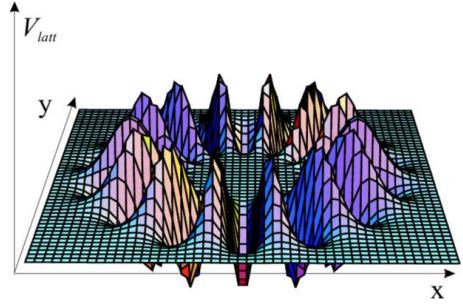} \\
\caption{The optical potential resulting from the interference of a
plane wave with an LG mode with $L=14$, $p=0$.
For $p\neq 0$ the potential is virtually unaltered}
\label{potenziale}
\end{figure}

So far open optical lattices have been studied. This constitutes a
limitation of optical apparata since a variety of studies for
finite 1d lattices with Periodic Boundary Conditions (PBC) exists
in the literature, that cannot be accesed with them.
In the same way as Gaussian laser beams are useful to produce open
optical lattices, we shall take advantage of the rotational symmetry of
 Laguerre--Gauss (LG) laser modes to produce closed optical lattices.
LG beams, obtained experimentally  making use of
computer generated holograms \cite{CHAVEZ},
have already been used in the field of  ultra-cold atoms\cite{ERTMER}.
A LG mode with frequency $\omega$, wave-vector $k$ and amplitude $E_0$
propagating along the $z$ axis can be written in cylindrical
coordinates $(r,\varphi,z)$ as \cite{Santamato}
$ 
E\left(r,\varphi\right)= E_0 f_{pl}(r) e^{il\varphi} e^{i(\omega
t
- k z)} $, 
$f_{pl}(r)={\displaystyle (-1)^p\sqrt{\frac{2p!}{\pi
\left(p+|l|\right)!}}\xi^{|l|} L_p^{|l|}\left(\xi^2\right)
e^{-\xi^2}}$,  $\xi=\sqrt{2}r/r_0 $,
where $r_0$ is the waist of the beam. and  $L_p^{|l|}$ are
associate Laguerre polynomials $L_m^n(x):= (-)^m d^m/dx^m
[L_{n+m}(x)]$, $L_{n+m}(x)$ being the Laguerre polynomials
themselves. The numbers $p$ and $l$ label the radial and azimuthal
quantum-coordinates, respectively.
The  lattice
modulation is obtained by interference of a LG beam with  a plane
wave $E_0e^{i(\omega t -kz)}$: in the far field, the interferogram
is periodic in $\varphi$ with $l$ wells. For even $l$ a perfect
$1d$-ring with $L=l$ lattice sites is obtained.
By
reflecting the combined beam (LG beam plus plane wave) back on
itself one achieves confinement also along $z$.
Indeed a series of disk shaped traps are obtained.
We point out that tunneling between the  disks $t_z$ can be made much weaker than
the corresponding tunneling within each ring $t_\phi$ adjusting $r_0/\lambda$ ({\it i. e.} focusing the LG beam). 
Such a  parameter depends monotonously only on $L$; for $L\gtrsim 15$,  $t_z/t_\phi \lesssim 10^{-2}$ can be achieved with $r_0/\lambda
\sim 100$.
The resulting lattice potential (see Fig.(\ref{potenziale})) is
described by
\begin{equation}
V_{latt}=4 E_0^2\left[1+ f^2_{pl} +2 f_{pl} \cos(l\varphi) \right ] \cos\left(k z\right)^2
\label{potondo}
\end{equation}
Note that, contrary to what was done in \cite{ERTMER},
here we need the laser
frequency to be tuned below the atomic resonance since we want to trap atoms into the ring.
For example with a laser intensity of $I=5W/cm^2$ and
$\Delta=-10^6MHz$  the potential wells would be separated by a
barrier of $\sim 5 \mu K$ much larger than the chemical potential
of a standard condensate (whose temperature can reach few $nK$);
with these parameters the scattering rate is $\ll 1$ photon/$sec$.
It is worth noting that, because of the relation:
$L_m^{|n-m|}(r)\leftrightarrow
H_n[(x-y)/\sqrt{2}]H_m[(x+y)/\sqrt{2}]$, LG modes can be  realized
also from Hermite--Gauss modes (modulo a $\pi/2$ phase change).
Such a ``mode-converter'', realized
experimentally in \cite{BEIJERSBERGEN}, 
can  switch  from an open  to a closed
lattice potential with the same periodicity and $L$. As we shall discuss further, this
device might be useful in the experiments.

We have just illustrated how to realize an optical lattice with
PBC. Now we show how to twist them.
The task
can be achieved by
applying an external, cone-shaped magnetic field ${\bf
B}=B_\varphi {\bf e}_\varphi+B_z {\bf e}_z$. In this way the
atomic magnetic dipoles ${\bf \mu}_{m_F}$ experience a field
varying along the ring, eventually equipping the periodic lattice
by a twist factor: $\Psi \rightarrow \displaystyle{e^{i
\phi_{m_F}}}\Psi $ at each winding, $\Psi$ being a generic  wave
function. The phase factor $\phi_{m_F}=m_F \pi \cos{\theta}$, with
$\tan{\theta}=B_\varphi/B_z$, is the analog of the Berry
phase\cite{NOTE} of the two state system corresponding to
the Zeeman splitting of the hyperfine atomic ground states; the role of time is  played by the angle $\varphi$. 
We can adjust  $\phi_{m_F}$ using an
additional laser beam (with a suitable frequency), relying on 
the AC-Stark shift: $A_E (m_F)$, where the function $A_E$ 
depends on the intensity of the laser and on the
Clebsch-Gordan coefficients 
corresponding to the matrix element of the electric dipole interaction energy\cite{AC-STARK}.
The resulting phase twist
is $\Phi_\sigma\doteq A_E (m_F)+ m_F\pi \cos \theta$  where $\sigma=m_F$.
Whereas boundary twists induced by a magnetic field pierceing the ring
are ``symmetric'', $\Phi_+ \equiv  \Phi_-$, our
protocol realizes $1d$-models with a  tunable $\Phi_\sigma$,
thus opening the
way to novel investigations discussed below.

\begin{figure}
\includegraphics[width=5.5cm]{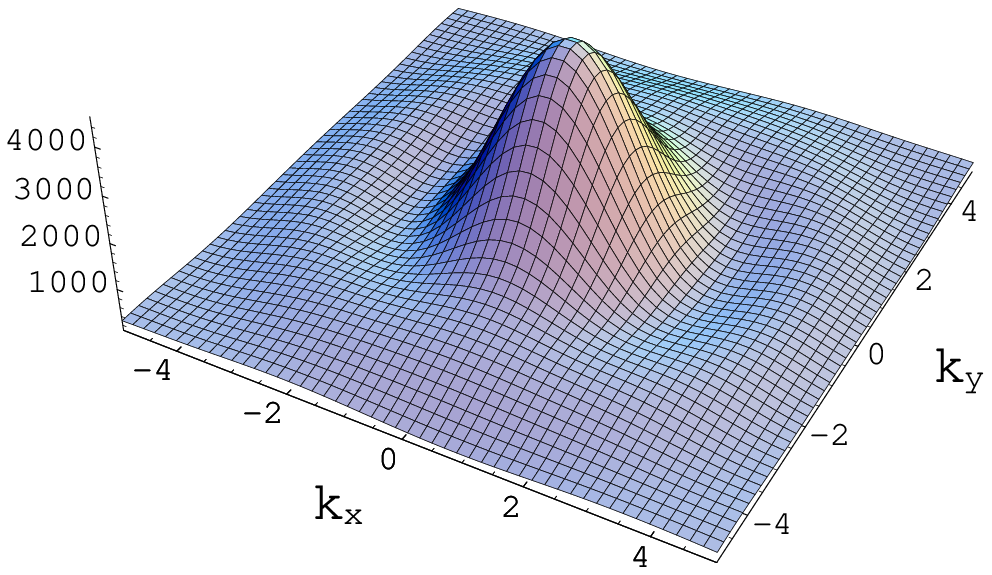}
\includegraphics[width=5.5cm]{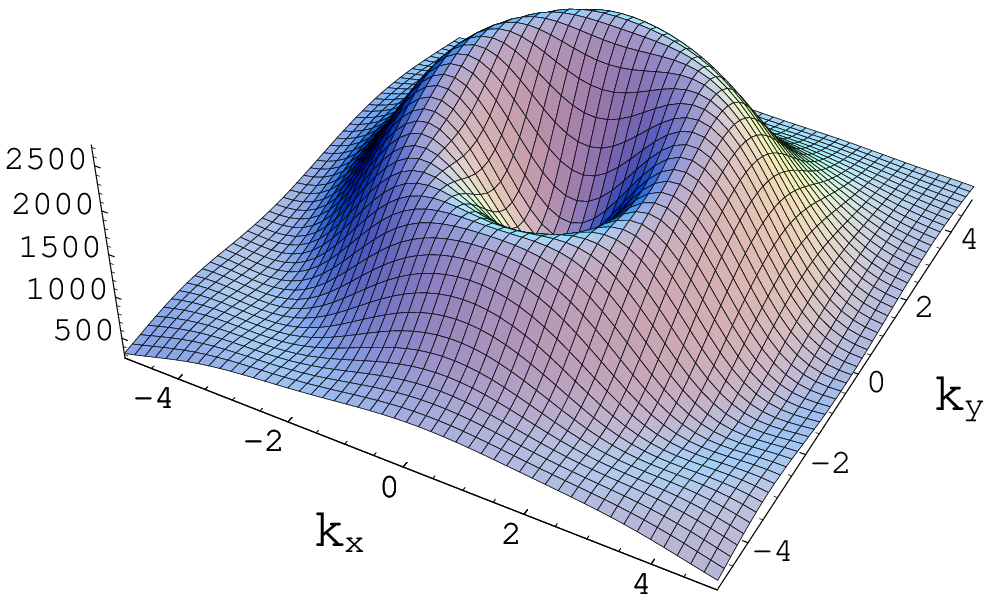} \\
\caption{The zero temp.
momentum distribution for
fermions with  Hubbard dynamics is presented: $|\Psi (k_x,k_y)|^2\propto |w(k_x,k_y)|^2 \sum_{i,j}e^{i {\bf k} \cdot ({\bf x}_i-{\bf x}_j)} \sum_{k_\phi}
e^{i k_\phi(\phi_i-\phi_j)} \langle n_{k_\phi} \rangle
$; ${\cal N}/L=32/16$
(${\cal N}/L=1$ is the less advantageous
case to discern the effects of $\Phi$ at finite size, since the
metal-insulator transition strongly suppresses $D_c$\cite{SCALAPINO});
$\langle n_{k_\phi} \rangle$ is
calculated perturbatively, at second order in $U/t$.
For $\Phi=0$ (left), $k_\phi= \{-\pi({\cal N}-1)/L\dots \pi
({\cal N}-1)/L\}$.  For $\Phi\neq 0$, $ k_\phi= \{-\pi ({\cal
N}-1)/L+\Phi/L \dots \pi ({\cal N}-1)/L+\Phi/L\}$; the asymmetry
in $|\Psi (k_x,k_y)|^2$ is due to the offset of $\langle
n_{k_\phi} \rangle$ caused by $\Phi$.}
\label{fourier}
\end{figure}

For OBC,  $\Phi_\sigma$  can be ``gauged away'' completely from the
system. In contrast, the boundary  phase cannot be eliminated for
closed loops and alters the phase diagram of the
system~\cite{SHASTRY-SUTHERLAND}. Infact $\Phi_\sigma$ emerges from the
sum of site dependent phases causing an increase of the velocity
field ($\propto$ to the tight binding amplitude $t$) that, in
absence of dissipation, may set a persistent current. Therefore
different regions in the phase diagram are identified depending on
the dynamical response of the system by perturbing  $\Phi_\sigma$.
The effect is reflected in the
curvature of the ${\cal N}$-particle energy levels $E_n$ respect to the
phase twist: $\rho_\sigma={L^2}\sum_n p_n \left
[E_n(\Phi_\sigma)-E_n(0)\right ]/({\cal N} t\Phi_\sigma^2) $,
where $p_n=e^{-\beta E_n }/Z$ are the Boltzmann weights.
For (spinless)
 bosons $\rho_+=\rho_-=\rho$ is proportional to the superfluid fraction.
Persistent currents  are studied analyzing   the charge stiffness
$D_c\propto \rho_+ +\rho_-$ (for electrons, it is the zero frequency
conductivity or Drude weight); a non vanishing $D_c$ sets a persistent current,
visible by releasing
the condensate for a time much longer than the typical atomic
oscillation period in the lattice wells.
\begin{figure}
\includegraphics[width=6cm]{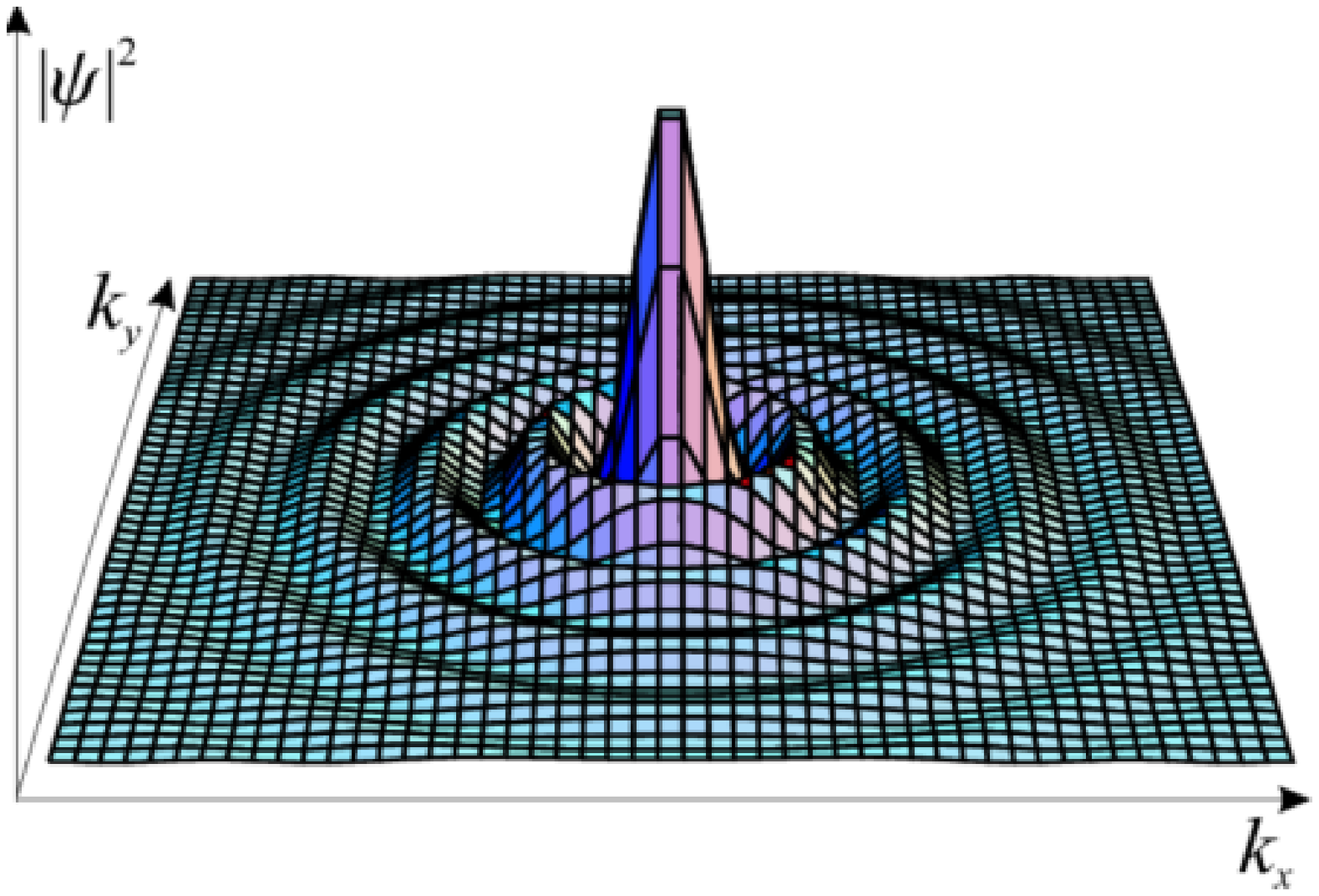} 
\includegraphics[width=6cm]{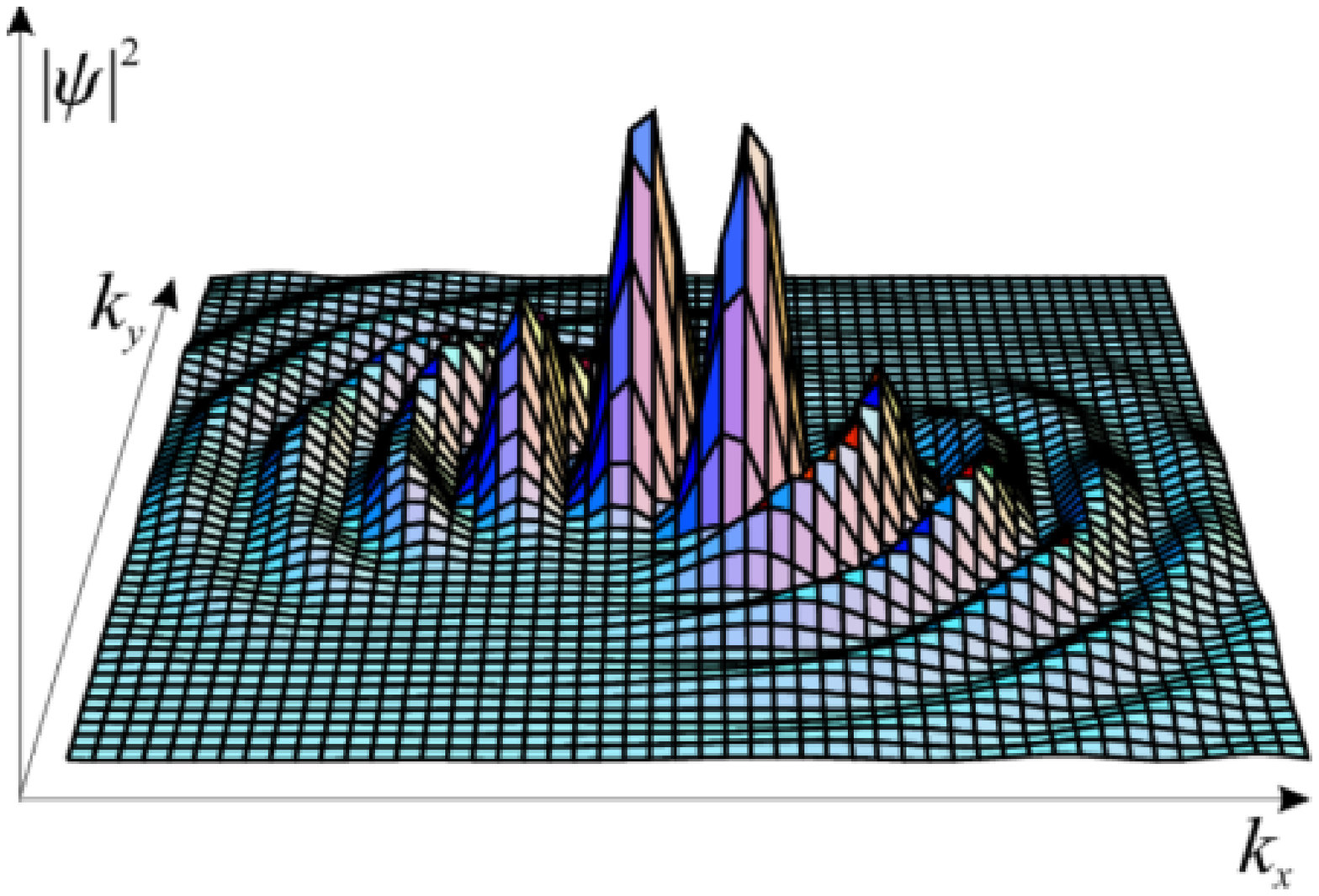}
\caption{Interference pattern for condensates released by
the lattice (\ref{potondo}), obtained resorting to the analog of light
diffraction from a circular grating\cite{PEDRI}. The figures
show the square of the order parameter $|\psi(k_x,k_y)|^2=|
\psi_0(k_x,k_y)\sum_{j=0}^{L-1} \cos\left [i (k_x\cos(2\pi j/L)+
k_y\sin(2\pi j/L) +\phi_j)\right ]|^2$. On the left  $\phi_j=0$ for all the
condensates; on the right the atoms move along the
ring with velocity $\propto \nabla \phi$;
the interference  pattern reflects a loss of matter at the
trap center caused by centrifugal effects.} \label{interf}
\end{figure}
Then the spatial distribution
of the condensates $ |\Psi({\bf r} ={\bf k}t,t)|$ is indicative of the initial
atomic momentum
distribution $|\Psi ({\bf k},0)|$\cite{PEDRI}; in particular
the phase difference between
atoms trapped in different sites, produces
characteristic interference patterns in the released condensates.
In  Fig.~(\ref{interf}) we  show such a
pattern for condensates released from the potential of
Fig.~(\ref{potenziale}) in mean-field approximation
(see also Fig.(\ref{fourier})).
Supercurrent/superfluid
fractions can be studied looking at the  response of the
system under imprinting of a dynamical phase
$\alpha_d(j,\sigma)\delta\tau \,$
to the atomic wave functions, flashing the atoms with an
additional Gaussian laser beam (can be much closer to
resonance than those creating the potential) with a waist larger
than the LG mode and with $\varphi$--dependent intensity.
The time $\delta\tau $ must be
too short to induce atomic motion by absorption during the pulse).

The case  $\Phi_+=-\Phi_-$  is useful to study the  spin stiffness
$D_s\propto\rho_+-\rho_-$ indicating long range spin correlations
in the system (for charged particles  $D_s$ would be proportional to the
inverse bulk spin susceptibility\cite{SHASTRY-SUTHERLAND}).
Generic values of $\Phi_+\neq \pm \Phi_- $
can be seen also as a result of certain correlated--hopping
processes (on the untwisted models)\cite{SCHULZ-SHASTRY}
and correspond to more exotic cases that, as far as we know,
have not been realized yet in physical systems.
To be specific  we consider ${\cal N}$ fermions described by the H
ubbard model with particle-density modulated kinetic energy
\begin{eqnarray}
\label{Hubbard}
H_{Hub} &=& -\sum_{j,\sigma} \mu_{j,\sigma} N_{{ j},\sigma} -  \sum_{j,\sigma }(
\tilde{t}_j (\sigma) c_{j+1,\sigma}^\dagger c_{j,\sigma} +h.c.) +U\, \sum_{ j}N_{{j},+} N_{{ j},-} \\
&\tilde{t}_j& (\sigma) =t \exp\Bigl[{\rm i}\gamma_j (\sigma)+ {\rm i} \sum_{l}^{}\bigl(\alpha_{j,l}(\sigma)N_{l,-\sigma}
   +  A_{j,l}(\sigma)N_{l,\sigma}\bigr)\Bigr] \;,
\label{correlated}
\end{eqnarray}
where $c_{j,\sigma}$'s are fermionic operators,  and $N_{l,\sigma}:= c_{l,\sigma}^\dagger c_{l,\sigma}$.
 $U=\pi b_s \int dx |w(x)|^4/m$ and
$t=\int dx w(x)[-\frac{1}{2m} \nabla^2+ V_{latt}] w(x+a)$,
($b_s$, $a$, and $w(x)$   indicates the scattering
length, the lattice spacing, and  Wannier functions respectively)
play the role of  the  Coulomb and hopping amplitudes respectively;
$\mu_{j,\sigma}$ is  of the order of the Bloch band separation\cite{SPIN};
the site dependence can be achieved by tuning the
magnetic confinement out of the symmetry axis of the optical ring.
For the model (\ref{Hubbard}) in a closed lattice, (\ref{correlated})
can be gauged away everywhere but at the boundary; therefore
(\ref{Hubbard}), (\ref{correlated}) is
equivalent to the ordinary Hubbard model, but with twisted
BC\cite{SCHULZ-SHASTRY}.
The phase twist is
$
\Phi_\sigma:=\phi(\sigma)+ \phi^{(1)}_{+-}(\sigma) {\cal
N}_{-\sigma} +\phi^{(1)}_{++}(\sigma)({\cal N}_\sigma-1) $,
where $
\phi^{(1)}_{+-}(\sigma)
= \sum_{j=1}^{L}\alpha_{j,m}(\sigma)\quad $, $\phi(\sigma)
=\sum_{j=1}^{L}\left (\gamma_j(\sigma)\;+\, A_{j,j}(\sigma)\right ) $,
$ \phi^{(1)}_{++}(\sigma)
=\sum_{{j=1\atop j\neq m-1,m}}^{L} A_{j,m}(\sigma)+A_{m,m-1}(\sigma)+A_{m-1,m+1}(\sigma)$.
  Hence, loading the Hubbard model into the twisted ring effectively
leads to the  physical realization of the model
(\ref{Hubbard}), (\ref{correlated}).
To point out the effects of $U$ (smearing of the Fermi distribution with
algebraic singularity at $k_F$)  in the persistent
current, $|\Psi ({\bf k})|^2$ is calculated for the
Hubbard ground state at small $U/t$,  and with 
$\Phi_\sigma=\phi(\sigma)$, $\phi(+)=\phi(-)=\phi$
(see  Fig.~(\ref{fourier})).

The proposed setups could
be used to study several issues in one dimensional systems.

I.\; The concept of conformal invariance plays a central role in
$1+1$ dimensional critical phenomena: universality is
characterized by a single parameter, the conformal anomaly $c$.
The physical meaning of $c$ resides in the concept of Casimir
energy, namely the variation of the vacuum energy density to  a
change in  the BC. For PBC it was
shown~\cite{CONFORMAL} that the finite size correction to the bulk
ground state energy is related to $c$: ${\cal
E}_{PBC}-{\cal E}_{bulk}=-\pi c v/6L$; resorting the modular
invariance this correction should be visible in the specific heat of the system, at low
temperature: $C(T)=\pi
c L T/3v $,  for each collective mode of the system; the speed of sound $v$ can
be extracted from the dispersion curve, at small $k$:
$v=\Delta {\cal E}/\Delta k$, for sufficiently large $L$ (for the
XXZ model, numerical analysis suggests that
 $L \gtrsim 15$\cite{BONNER}).
Except for integrable models, it is hard to measure or even have
numerical estimates of $c$ in solid state
systems\cite{EXPERIMENTAL-c,CONFORMAL}. With the presented setups
for highly controllable loaded models these
measurements can be done with unique accuracy. Both $C(T)$ and
$\Delta {\cal E}/\Delta k$ can be  measured following the
techniques employed by Cornell {\it et al.}\cite{JIN}. To discern
finite size effects in $C(T)$ the PBC to OBC converter,
discussed above, could be a valid tool. Indeed, the finite
size correction to ${\cal E}_{bulk}$ for OBC is also
proportional to $c$, but with a {\it different}
coefficient\cite{CONFORMAL}. Then: $
{\displaystyle c v=\frac{8 L}{\pi} ({\cal E}_{PBC}-{\cal E}_{OBC})+ {\cal F}_S}
$,
where  ${\cal F}_S$ is
the bulk limit of the surface energy that, being  non-universal,
can be fixed by performing the measurements for different $L$.
(mimicking a ``finite size scaling analysis'').
Remarkably, both the   energies  ${\cal E}_{PBC}$ and  ${\cal E}_{OBC}$ might
be accessible measuring the second moment of the velocity of the released
condensate\cite{GROUND}.

II.\; A general model we can engineer in the ring shaped lattice is
\begin{equation}
\label{mixture}
H=H_{BH}+H_{Hub}+H_I
\end{equation}
where $H_{BH}$ is the Bose-Hubbard Hamiltonian 
\cite{GREINER} and $H_I$ describes a density-density,
fermion-boson  interaction\cite{MIXED}. By tuning $\Delta$ within
the  fine structure of  the fermionic atoms, a spin dependence can
be inserted in the hopping amplitude   of the Hubbard model:
$t\rightarrow t_\sigma$. At $ {\cal N}/L=1$ and $t_\sigma\ll U $ the
Hubbard ring effectively accounts for the physical realization of
the twisted $XXZ$ model with  anisotropy $\gamma=(t_+^2+t_-^2)/(2
t_+ t_-) $ and external field $h=4 \sum_\sigma \sigma
t_\sigma^2/\mu_\sigma$\cite{SPIN}. Loading quantum systems
described by Hamiltonians of the type (\ref{mixture}) in lattices
with twisted BC could serve to study
charge and spin stiffness in physical systems  with  tunable
interaction and/or disorder. For example  a mixture of $^{87} Rb$
and $^{40} K$ atoms constitutes an ideal system to check
the recent experimental evidence suggesting that the supersolid
order\cite{SARO-REV} would be effectively favoured by the
insertion of fermionic degrees of freedom into homogenous bosonic
systems. The off-diagonal long range order manifests in
superfluid currents. Jumps between
non-vanishing supercurrents should reveal the existence of the
supersolid phase\cite{SS-HELIUM}. This  should be
accompanied by a macroscopic occupation in the condensate at
a non vanishing wave vector ($\sim \pi/(na)$, $n\ge 2$) signalling the
charge-density-wave instability\cite{BLATTER}. 
The two condensates should be traced in the interference fringes.

It was proved that  exactly solvable twisted Hubbard/$XXZ$
rings~\cite{SHASTRY-SUTHERLAND,OSTERLOH-DEF} are equivalent to
untwisted models for particles with intermediate statistics; this
results in  modifications of the exponents of the (low energy)
correlation functions\cite{SCHULZ-SHASTRY}. The spatial profile of
the latter  might be detected by photoassociation   techniques, as
suggested in \cite{GARCIA}.

III\; Another  interesting issue we
can study  is the conjecture\cite{CASTELLA-PRE}
that Poisson $or$ Wigner-Dyson level-statistics manifest in that the
thermal Drude weights have qualitatively different slopes for
integrable (smooth algebraic temperature-decrease, universal
behaviour of $D(T)/D(0)$) $or$ non-integrable (sharp,
non-universal suppression of  $D(T)/D(0)$) systems. Due to the
precise knowledge of the model-Hamiltonian  under analysis    we
can address the problem directly in a physical system. For
example, we could consider $^{40}K$ pure-$XXZ$ rings with twisted
BC, for different $L$'s; using the Feshbach resonance one could tune
$b_s\sim 2a$; the resulting $XXZ$  model with
next-nearest neighbor density-density interaction is
non-integrable (another way is to destroy the integrability
introducing disorder into the ring by site-dependent $h_j$). In
short: integrability can be switched on and off by tuning the
Feshbach resonance (or adjusting the energy offsets $h_j$).
The presence of persistent currents can be detected
 along the lines described above (see Fig.(\ref{interf}), (\ref{fourier})).
Numerical investigations for the $XXZ$ model suggest that the effect should be visible for $ T/L\gtrsim 0.1 \gamma $~\cite{CASTELLA-ZOTOS}.

In summary we have suggested a number of protocols
to realize closed  rings of many quantum particles,
by optical means. This is possible by employing
slight variations and combinations of techniques already developed within
the current experimental activity in atomic physics.
We have discussed how several open problems in condensed matter
physics can be enlightened by such a setup.
We finally observe that the clockwise/anticlockwise currents in a
few-wells-ring  constitutes a controllable two state-system
analogous to the flux-qubit realized by a SQUID. As the
current is neutral, the corresponding
decoherence-rate is much lower compared to solid state devices
(charged currents). Information transfer could be
mediated by an induced-dipole--dipole
atomic interaction.

{\bf Acknowledgments}.
We thank A. Cappelli, M. Inguscio, G. Falci, R. Fazio, and H. Frahm for
support and discussions.

\end{document}